\PassOptionsToPackage{unicode}{hyperref}
\PassOptionsToPackage{hyphens}{url}
\PassOptionsToPackage{dvipsnames,svgnames,x11names}{xcolor}
\documentclass[
  10pt,
]{article}
\usepackage{amsmath,amssymb}
\usepackage{iftex}
\ifPDFTeX
  \usepackage[T1]{fontenc}
  \usepackage[utf8]{inputenc}
  \usepackage{textcomp} 
\else 
  \usepackage{unicode-math} 
  \defaultfontfeatures{Scale=MatchLowercase}
  \defaultfontfeatures[\rmfamily]{Ligatures=TeX,Scale=1}
\fi
\usepackage{lmodern}
\ifPDFTeX\else
\fi
\IfFileExists{upquote.sty}{\usepackage{upquote}}{}
\IfFileExists{microtype.sty}{
  \usepackage[]{microtype}
  \UseMicrotypeSet[protrusion]{basicmath} 
}{}
\makeatletter
\@ifundefined{KOMAClassName}{
  \IfFileExists{parskip.sty}{%
    \usepackage{parskip}
  }{
    \setlength{\parindent}{0pt}
    \setlength{\parskip}{6pt plus 2pt minus 1pt}}
}{
  \KOMAoptions{parskip=half}}
\makeatother
\usepackage{xcolor}
\usepackage[margin=0.9in]{geometry}
\usepackage{color}
\usepackage{fancyvrb}

\DefineVerbatimEnvironment{Highlighting}{Verbatim}{commandchars=\\\{\}}
\newenvironment{Shaded}{}{}

\newcommand{\AttributeTok}[1]{\textcolor[rgb]{0.49,0.56,0.16}{#1}}

\newcommand{\ExtensionTok}[1]{#1}

\newcommand{\NormalTok}[1]{#1}

\usepackage{longtable,booktabs,array}
\usepackage{calc} 
\usepackage{etoolbox}
\makeatletter
\patchcmd\longtable{\par}{\if@noskipsec\mbox{}\fi\par}{}{}
\makeatother
\IfFileExists{footnotehyper.sty}{\usepackage{footnotehyper}}{\usepackage{footnote}}
\makesavenoteenv{longtable}
\usepackage{graphicx}
\makeatletter
\def\maxwidth{\ifdim\Gin@nat@width>\linewidth\linewidth\else\Gin@nat@width\fi}
\def\maxheight{\ifdim\Gin@nat@height>\textheight\textheight\else\Gin@nat@height\fi}
\makeatother
\setkeys{Gin}{width=\maxwidth,height=\maxheight,keepaspectratio}
\makeatletter
\def\fps@figure{htbp}
\makeatother
\providecommand{\tightlist}{%
  \setlength{\itemsep}{0pt}\setlength{\parskip}{0pt}}
\usepackage{amsmath,amssymb,booktabs}
\usepackage{float,needspace}
\ifLuaTeX
  \usepackage{selnolig}  
\fi
\usepackage{bookmark}
\IfFileExists{xurl.sty}{\usepackage{xurl}}{} 
\hypersetup{
  pdftitle={Formalizing Cooperation and Competition: An AI-Assisted Mathematical Addendum to a 2018 Planning Thesis},
  pdfauthor={Haitham Masaud Amar},
  colorlinks=true,
  linkcolor={Maroon},
  filecolor={Maroon},
  citecolor={Blue},
  urlcolor={Blue},
  pdfcreator={LaTeX via pandoc}}

\title{Formalizing Cooperation and Competition: An AI-Assisted
Mathematical Addendum to a 2018 Planning Thesis}
\author{Haitham Masaud Amar}
\date{21 October 2026 --- preprint draft}

\begin{document}
\maketitle
\begin{abstract}
We present an AI-assisted mathematical addendum to Strategic Trip
Planning: Striking a Balance Between Competition and Cooperation (Amar,
2018). We reconstruct selected relationships between strategic choice,
cooperative stability, participant approval and enforcement. The
thesis's Nash-to-core proposition is under-specified until outsider
opportunities are stated. We verify two-person implications under fixed
outsiders and explicit efficiency conditions. Under isolation, the
original example has an empty TU core; its funded counterpart uses an
explicit external subsidy. For a monotone resource model we characterize
the exact financing gap. A finite negotiation procedure implemented in
Lean is proved to terminate with participant approval, joint-cost
non-worsening and local optimality against permitted approved moves.
Supporting formalizations address bargaining geometry and regret-based
efficiency. A resource-game counterinstance shows why monotonicity alone
provides no uniform price-of-anarchy bound. Reproducible computations
compare a documented heuristic reconstruction with the verified
procedure on 44 source-based instances and 96 starting cases from four
additional synthetic network families. Acceptance-matched heuristic and
randomized pairwise-search controls contextualize the optimization
results. The mathematical ingredients are established results; the
contribution is a traceable reconstruction linking source statements,
explicit assumptions, checked proofs, executable code and retained
counterinstances. The study illustrates how AI-assisted investigation
can turn engineering intuitions from legacy research into precise
mathematical claims whose conditions and computational consequences can
be inspected and reproduced.
\end{abstract}

\section{1. Introduction}\label{introduction}

Trip planning is individually motivated but jointly realized. A route
that is attractive when considered in isolation may become costly once
other travellers select overlapping resources. Conversely, a traveller
may accept a less-preferred route because it reduces conflict, improves
a joint outcome or forms part of an enforceable agreement. These
observations motivated my doctoral framework, \emph{Strategic Trip
Planning: Striking a Balance Between Competition and Cooperation}
{[}1{]}. Its principal components were personalized route generation,
team formation and bargaining, with a later application to territory
sharing in social taxi networks.

My decision to revisit the thesis was inspired by OpenAI's 6 October
2026 release of mathematical manuscripts and supporting proof artifacts,
including Lean formalizations for many of the proofs {[}10{]}. That
release prompted a personal question: could AI-assisted mathematical
investigation help me give the engineering intuitions developed in my
2018 doctoral work a fuller and more precise mathematical treatment? The
thesis's mathematics had supported an engineering contribution; it had
not been developed as a separate contribution to mathematical theory. I
wanted to investigate how far its underlying ideas could be
reconstructed as explicit definitions, justified assumptions and
checkable theorems, and how the resulting treatment would behave on new
synthetic planning instances.

This motivation shapes the present work. Lean substantiates the formal
guarantees of the stated models and revised procedure, while new
reproducible simulations test their engineering behavior and compare
outcomes with exact finite benchmarks. Together, these methods give a
concrete route from engineering intuition to a mathematical formulation
whose assumptions, proofs and computational consequences can be
inspected. The aim is to substantiate the coordination outcomes that
motivated the original work and establish the conditions under which
they follow. The simulations provide finite-instance evidence; universal
guarantees come from the theorems and their explicit assumptions.

This paper is an addendum to my thesis and a case study in AI-assisted
reconstruction of legacy engineering mathematics. Its purpose is to show
which mathematical statements can be made precise, which assumptions
change their meaning, and how the resulting claims can be checked
against executable procedures. The principal result of the investigation
is the source-to-model-to-proof correspondence, together with the
evidence needed to assess its limits.

The starting point remains the observation in the thesis abstract:
``Mobility planning as it stands today is an individual decision-making
effort that takes place in an environment governed by the collective
actions of various competing travellers'' {[}1, p.iv, PDF p.4{]}. The
framework accordingly accommodates ``the travellers' individual
behaviours, preferences, and goals in cooperative and competitive
scenarios'' {[}1, p.iv, PDF p.4{]}. These statements motivate the models
developed here; they are not offered as claims about the present state
of transportation research.

The engineering objective is coordination among travellers who retain
their own agendas, rather than the optimization of a centrally
controlled fleet {[}1, Chapter 1, pp.3--4, PDF pp.23--24{]}. In the
original framework, traveller-centric trip planning supplies
personalized strategies, team formation determines who interacts, and
bargaining manages conflicts. The present paper develops the
mathematical structure underlying that architecture. It does not repeat
the original experimental study or require the original engineering
contribution to have been a contribution to mathematical theory.

This paper develops those ideas as a collection of compatible
mathematical models. Our aim is not to identify cooperation with one
equilibrium concept. We instead ask what must be specified for the
relationships between competitive choice, cooperative stability and
collective efficiency to hold. Four questions guide the treatment: what
actions and costs define the strategic game; what opportunities remain
available to a departing coalition; what the referee actually enforces;
and which costs a repeated-game learner minimizes.

The distinction concerning departing coalitions is decisive. A
unilateral Nash deviation holds opponents' actions fixed. A cooperative
opportunity may hold outsiders fixed, remove them, allow them to respond
strategically or allow the deviating coalition to escape a pricing
mechanism. These conventions cannot be interchanged. Once they are made
explicit, several intended connections become provable. Other
connections require additional conditions or fail under a different,
equally well-defined convention.

The treatment connects three contributions. First, it specifies
coalition opportunities and financing so that strategic and cooperative
stability claims have precise meanings. Second, it implements a finite
negotiation procedure and proves acceptance, termination and the
resulting local improvement condition. Third, it compares a documented
reconstruction of the thesis heuristic with that executable procedure on
shared finite instances. The original two-traveller model supplies an
end-to-end example whose returned plans also satisfy the stronger
equilibrium and cooperative conditions. The enforcement and regret
results explain what additional mechanisms can certify about
implementation and repeated outcomes.

The new synthetic data are a substantive part of this contribution. The
thesis supplies the planning concepts, source examples and heuristic
structure; the present treatment makes the relevant assumptions
explicit, proves a specified coordination procedure and tests
implementations on newly generated inputs. This provides fresh evidence
that the coordination logic can operate beyond the original numerical
example. Reproducible generators, shared-instance comparisons and
exhaustive finite benchmarks allow that evidence to be inspected and
rerun. The resulting support concerns the engineering framework under
the stated reconstruction and model assumptions; the guarantees of the
revised procedure are identified separately from observations about the
reconstructed heuristic.

The formal development uses Lean 4.24.0 {[}8{]} and Mathlib v4.24.0
{[}9{]}. We retain an assumption ledger and an exact correspondence
between the planning concepts and formal declarations. The present
treatment specifies the counterpart rules and admissible financing
explicitly. These specifications are part of the revised model, whose
guarantees are proved below; the original thesis is preserved unchanged.

\section{2. Strategic actions and cooperative
opportunities}\label{strategic-actions-and-cooperative-opportunities}

Let \(N\) be a finite nonempty player set. Player \(i\) has a finite
nonempty action set \(A_i\), and \(A=\prod_i A_i\) is the profile space.
Actions may represent routes, territories or explicitly discretized
departure choices. We write \(c_i(s)\) for player \(i\)'s physical cost
and

\[C(s)=\sum_{i\in N}c_i(s).\]

Here ``physical cost'' means the underlying unpriced planning cost,
distinguished from referee charges and observed learning scores. Its
interpretation may be travel cost or a declared preference score; the
formal model does not turn a subjective score into a measured travel
time. Summing personalized scores requires a declared common scale and
system objective. We assume that scale where aggregate costs or TU
payments are used.

The principal notation is:

\begin{longtable}[]{@{}
  >{\raggedright\arraybackslash}p{(\columnwidth - 2\tabcolsep) * \real{0.5000}}
  >{\raggedright\arraybackslash}p{(\columnwidth - 2\tabcolsep) * \real{0.5000}}@{}}
\toprule\noalign{}
\begin{minipage}[b]{\linewidth}\raggedright
Symbol
\end{minipage} & \begin{minipage}[b]{\linewidth}\raggedright
Meaning
\end{minipage} \\
\midrule\noalign{}
\endhead
\bottomrule\noalign{}
\endlastfoot
\(s\), \(b\), \(s^*\) & Current profile, feasible baseline, joint
comparator \\
\(c_i(s)\), \(C(s)\) & Base player cost and its declared sum \\
\(v_s(S)\), \(v_{\mathrm{iso}}(S)\) & Fixed-outsider and isolated
coalition opportunities \\
\(h_i\), \(F_h\) & Participant acceptance ceiling and approved
profiles \\
\(\widehat c_i\), \(L_i^t\) & Enforced cost and time-dependent observed
cost \\
\(R_i(T)\), \(B_i(T)\) & All-action regret bound and cumulative
cost-distortion bound \\
\end{longtable}

The full-profile argument records resource externalities. A baseline
route score on \(A_i\) alone does not specify them. A Nash equilibrium
satisfies

\[c_i(s)\le c_i(a_i,s_{-i})\qquad(\forall i,\ a_i\in A_i). \tag{1}\]

A social optimum minimizes \(C\) over feasible profiles. Pareto
efficiency means that no profile makes every player weakly better and at
least one player strictly better. These are separate properties. We do
not assume that independently choosing baseline favourites produces a
Nash equilibrium.

The source supplies finite nonempty action sets in Chapter 3, pp.39--42
(PDF pp.59--62), and distinguishes personalized and system utilities in
Chapter 5, pp.80--81. Its use of compact convex strategy sets in Chapter
4, pp.55--56 requires a representation beyond a finite list of distinct
deterministic routes. Finite probability distributions provide one
explicit representation. Independent product distributions describe
mixed strategies; arbitrary joint distributions allow correlation. No
deterministic trip constraint is silently replaced by an expected
constraint.

\subsection{2.1 Transferable and nontransferable
stability}\label{transferable-and-nontransferable-stability}

For a scalar coalition-cost function \(v:2^N\to\mathbb R\) with
\(v(\varnothing)=0\), a TU cost-core allocation satisfies

\[x(N)=v(N),\qquad x(S)\le v(S)\quad(\forall S\subseteq N), \tag{2}\]

where \(x(S)=\sum_{i\in S}x_i\). One fixed allocation is used for every
coalition. We follow the source's cost orientation in (4.23), p.65, and
(6.3), p.126. Its earlier reward orientation reverses the inequalities.
Different personalized utilities alone do not justify transferable
utility. TU conclusions below concern a declared common cost/payment
scale and permit real transfers.

An NTU model instead retains attainable cost vectors. A nonempty
coalition blocks an attainable vector if it has another attainable
vector that strictly lowers every member's cost. This formulation uses
no interpersonal transfer. A coalition's aggregate cost may decrease
while one member loses; that is a potential TU improvement, but not
necessarily an NTU block.

\subsection{2.2 Two explicit outsider
conventions}\label{two-explicit-outsider-conventions}

Fix a reference agreement \(s\). The fixed-outsider value is

\[v_s(S)=\min_{t:\ t_{-S}=s_{-S}}\sum_{i\in S}c_i(t). \tag{3}\]

The admissible set is nonempty because it contains \(s\), and a minimum
exists in the finite game. For NTU opportunities, retain the vectors
\((c_i(t))_{i\in S}\) under the same constraint. Equation (3) defines a
reference-dependent family of cooperative problems; it is not a
reference-independent characteristic function.

Under isolation, outsiders are removed from the resource environment and
coalition costs are recomputed. We denote the resulting minimum by
\(v_{\mathrm{iso}}(S)\). This convention needs an explicit
coalition-level resource model, supplied in Section 4. A coalition that
leaves the referee's jurisdiction may have yet another opportunity. Exit
rights are therefore part of a mechanism specification.

\section{3. A precise equilibrium-to-core
bridge}\label{a-precise-equilibrium-to-core-bridge}

Proposition 4.2.1 of the planning framework {[}1, p.52, PDF p.72{]}
connects strategic stability with cooperative stability. The source does
not uniquely specify the outsider opportunities needed to determine that
connection. We formulate two conditional counterparts: fixed-outsider
opportunities for two-person agreements, and financed isolation for
efficient Nash agreements in the monotone resource model. The following
characterization supplies the first proof.

\textbf{Theorem 1 --- Fixed-outsider characterization.} Under (3), the
realized cost allocation \((c_i(s))_i\) belongs to the TU cost core if
and only if

\[\sum_{i\in S}c_i(s)\le\sum_{i\in S}c_i(t)
\quad\text{whenever }t_{-S}=s_{-S}. \tag{4}\]

For NTU attainable vectors under the same outsider rule, the realized
vector belongs to the NTU core if and only if \(s\) is strong Nash: no
nonempty coalition can change its actions, holding outsiders fixed, and
strictly improve every member.

\emph{Proof.} A TU core inequality bounds the current coalition cost by
its minimum opportunity, and therefore by every admissible deviation.
Conversely, (4) bounds the current cost by an attained minimum. The
reference profile itself is admissible, giving the opposite inequality.
Applying this to \(N\) establishes efficiency. For NTU, admissible
blocking vectors and improving joint deviations have exactly the same
witnesses. The TU and NTU conclusions use different improvement tests.
\(\square\)

\textbf{Theorem 2 (Proposition 4.2.1A) --- Two-traveller fixed-outsider
counterpart.} For two players:

\begin{enumerate}
\def\labelenumi{\arabic{enumi}.}
\tightlist
\item
  Nash equilibrium plus Pareto efficiency implies the fixed-outsider NTU
  core.
\item
  Nash equilibrium plus social optimality implies the fixed-outsider TU
  cost core.
\end{enumerate}

\emph{Proof.} The only nonempty coalitions are the two singletons and
the grand coalition. Nash rules out singleton blocks. Pareto efficiency
rules out a grand-coalition strict improvement for both members. For TU,
social optimality rules out a reduction of the grand coalition's sum
cost. This establishes the corresponding conditions in Theorem 1.
\(\square\)

The second premise in each statement is essential to the argument. For
more than two players there are intermediate coalitions, so this proof
does not extend merely because a negotiation procedure uses pairs. The
result is a sufficient completion of the two-person theory, not a proof
that arbitrary pairwise bargaining produces a globally stable agreement.

\subsection{3.1 The original two-traveller
instance}\label{the-original-two-traveller-instance}

Table 4.1, p.59 (PDF p.79), becomes the following positive-cost table
after negating the displayed payoffs:

\begin{longtable}[]{@{}lrr@{}}
\toprule\noalign{}
Traveller 1 / Traveller 2 & \(p_1\) & \(p_2\) \\
\midrule\noalign{}
\endhead
\bottomrule\noalign{}
\endlastfoot
\(p_1\) & \((8,8)\) & \((4,5)\) \\
\(p_2\) & \((5,4)\) & \((7,7)\) \\
\end{longtable}

The relevant Figure 4.2 routes have private prefixes of cost 2. The
first action uses shared suffix edges \(C\to B\) and \(B\to t\), with
base cost 1 each; the second uses \(C\to t\), of base cost 3. Each
shared suffix edge adds 2 per traveller. The formal resource calculation
reproduces every table entry. We use the two actions explicitly selected
in equations (4.2)--(4.5), rather than adding routes not available in
this example.

At \((p_1,p_2)\), the unilateral alternatives cost 7 instead of 4 and 8
instead of 5. This profile is Nash. Its total cost is 9, the minimum
among all four profiles, and it is Pareto efficient. Thus both
conclusions of Theorem 2 apply. The source's two-person compromise
intuition has a rigorous positive realization under fixed-outsider
opportunities.

Under isolation, each singleton achieves cost 4, while both travellers
together cost at least 9. Equation (2) would require \(x_1+x_2=9\) and
\(x_1,x_2\le4\), which is impossible. Every correlated lottery still has
expected total cost at least 9, so randomization does not remove this
obstruction. The illustrative example remains valid as an explanation of
cooperation; it exposes why the cooperative opportunity convention must
also be specified.

The example therefore fixes the engineering interpretation of the
two-person result: a jointly efficient strategic compromise is
cooperatively stable when departure opportunities retain the other
traveller's reference plan. If departure removes the other traveller
entirely, the financing must also be specified. Section 4.2 supplies
that specification and its proof.

\subsection{3.2 Bargaining geometry of the two-traveller
example}\label{bargaining-geometry-of-the-two-traveller-example}

Figure 1 expresses the same example in utility gains relative to the
declared status-quo costs \((8,8)\): \(g_i=8-\mathbb{E}[c_i]\). The four
deterministic gain vectors are \((0,0)\), \((4,3)\), \((3,4)\) and
\((1,1)\). Permitting joint lotteries gives their convex hull. The
segment joining \((4,3)\) and \((3,4)\) is the efficient expected-payoff
frontier, on which total expected cost is 9. Its endpoints are the two
deterministic pure Nash equilibria; neither equilibrium selects a unique
division of the gains. The chosen status quo is an illustrative planning
baseline, not a derived noncooperative threat outcome.

For this declared disagreement point and expected-utility
representation, the classical Nash bargaining product {[}7{]} is
\(g_1g_2\). We verify in Lean that every joint distribution satisfies
\(g_i\ge0\), \(g_1+g_2\le7\) and \(g_1g_2\le49/4\). A half-and-half
lottery between the two split profiles has expected costs \((9/2,9/2)\)
and gains \((7/2,7/2)\), attaining the bound. The dashed product curve
is tangent to the efficient segment at that point. This provides a
precise geometric interpretation of equal expected gains; it is an
explicit bargaining construction, not a claim that Algorithms 1--2
optimize the Nash product. The public joint lottery is distinct from
independent mixed strategies and from a deterministic route assignment.
No hard constraint is replaced by an expected constraint in the planning
procedure.

\begin{figure}
\centering
\includegraphics[width=4.8in,height=\textheight]{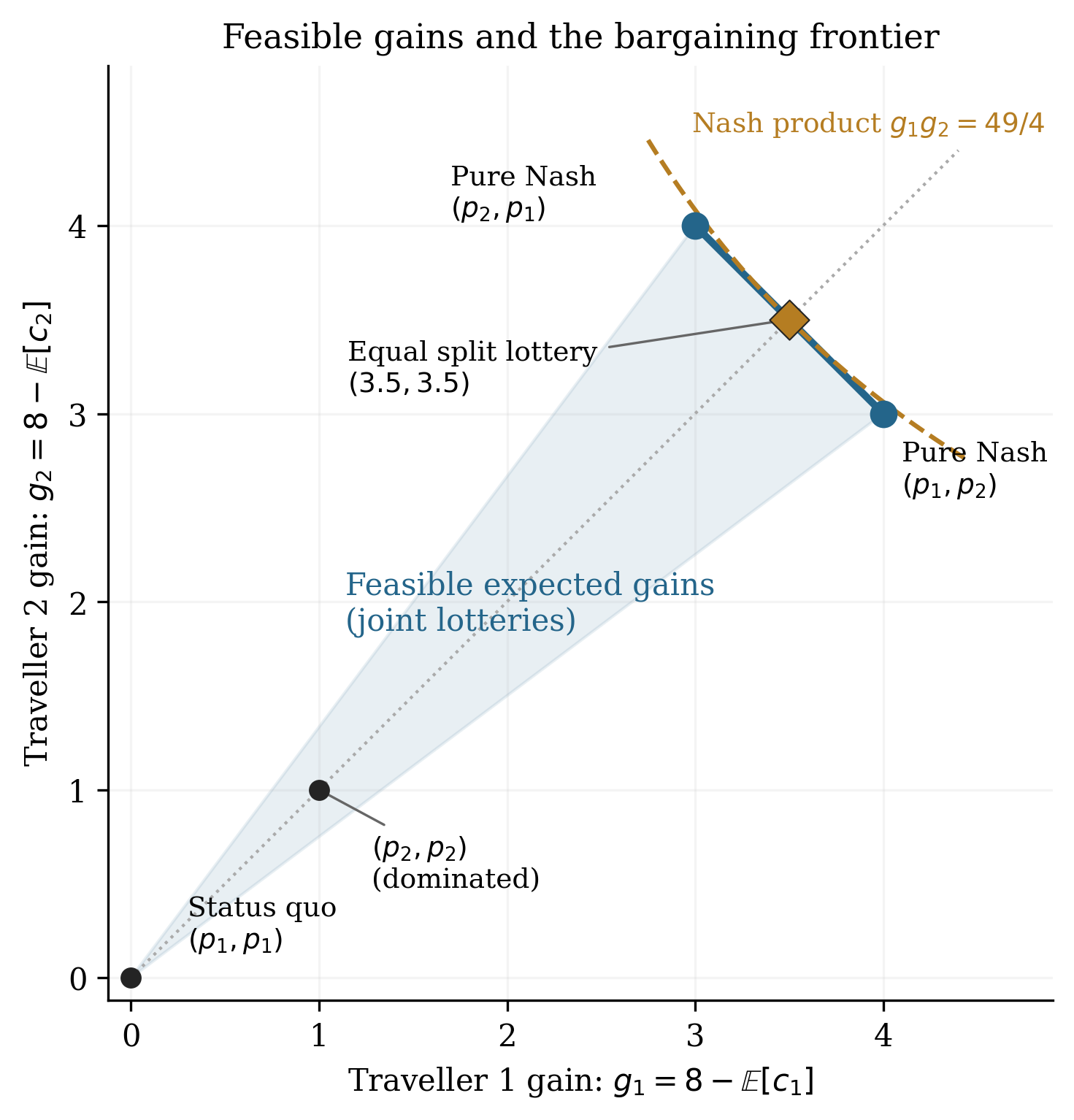}
\caption{Feasible expected gains, the efficient frontier, pure Nash
outcomes and an attainable Nash-product maximizer. Gains use the
interacting status quo \((8,8)\), not isolated exit costs. The shaded
triangle represents joint lotteries over the displayed deterministic
outcomes.}
\end{figure}

\subsection{3.3 Adopted counterpart rule for the revised
proposition}\label{adopted-counterpart-rule-for-the-revised-proposition}

For the revised two-traveller formulation, we adopt fixed-outsider
opportunities (3) as the counterpart rule. Travellers retain their
personalized feasible actions; a departing coalition may coordinate its
members' actions, but nonmembers continue their reference plans. We use
the NTU version when costs cannot be transferred. Under this convention,
the replacement for Proposition 4.2.1 is Theorem 2: a Nash agreement
that is Pareto efficient has a nonempty fixed-outsider NTU core; a Nash
agreement minimizing joint cost has the corresponding TU core. The
source example satisfies both conditions. These rules define Proposition
4.2.1A in the present treatment, and both conclusions hold for the
original example.

Proposition 4.2.1A is thus resolved under a fully defined counterpart
rule and explicit efficiency assumptions. Its scope is two-person
agreements; the multi-traveller financed result is given in Proposition
4.2.1B. Core nonemptiness is the cooperative conclusion proved here. The
classical balanced-collection characterization is not an additional
machine-verified theorem of this paper.

\section{4. Congestion, coalition budgets and the cost of
stability}\label{congestion-coalition-budgets-and-the-cost-of-stability}

We now give isolation a complete resource interpretation. Each action
\(a_i\) uses a finite set \(U_i(a_i)\) of resources. It has a private
cost \(b_i(a_i)\), and resource \(e\) has a common nondecreasing
congestion cost \(\ell_e(k)\). Define

\[n_e(S,t)=|\{i\in S:e\in U_i(t_i)\}|,\]

\[c_i^S(t)=b_i(t_i)+\sum_{e\in U_i(t_i)}\ell_e(n_e(S,t)),\qquad
v(S)=\min_{t\in A}\sum_{i\in S}c_i^S(t). \tag{5}\]

Outside actions do not affect (5), although a full profile conveniently
indexes coalition choices. Finite nonempty action sets give attained
minima for every coalition. The physical game uses \(c_i=c_i^N\).
Nonnegative private and resource costs match the transportation
interpretation; the inequalities below only need monotonicity.
Player-specific private costs and asymmetric feasible routes are
allowed.

This is a source-supported subclass of the models in Chapters 3, 4 and
6: it retains additive resource conflicts and nondecreasing
concentration costs. It does not represent every possible reward map
\(\zeta\), coalition production benefit, discount or private fare
mechanism. Those require a different model and may reverse the budget
conclusion.

\textbf{Theorem 3 --- Singleton domination from congestion.} The
isolated values in (5) satisfy

\[v(S)\ge\sum_{i\in S}v(\{i\}). \tag{6}\]

\emph{Proof.} Whenever \(i\in S\) uses resource \(e\), its load is at
least one. Monotonicity implies \(\ell_e(n_e(S,t))\ge\ell_e(1)\). Thus
its coalition cost is at least its solo cost for the same action, which
is at least its optimized singleton cost. Summing over \(S\) and
evaluating at an attained coalition minimum proves (6). The identity
between singleton coalition cost and solo action cost is also verified
separately. \(\square\)

Call any cost game satisfying (6) \emph{singleton dominated}. Set

\[G=v(N)-\sum_i v(\{i\})\ge0. \tag{7}\]

The gap \(G\) measures the additional simultaneous-service cost relative
to isolated singleton promises. It is not a Nash inefficiency measure:
it may be positive even at a social optimum.

\textbf{Theorem 4 --- Exact isolated-core criterion.} For a
singleton-dominated cost game,

\[\operatorname{Core}(v)\ne\varnothing\quad\Longleftrightarrow\quad G=0. \tag{8}\]

\emph{Proof.} Singleton core inequalities imply
\(v(N)=x(N)\le\sum_i v(\{i\})\). Equation (6) supplies the reverse
inequality. If equality holds, \(x_i=v(\{i\})\) is efficient and
satisfies every coalition bound by (6). \(\square\)

This result identifies the actual obstruction. In this class, a payment
rearrangement cannot cover a positive total gap while guaranteeing every
isolated singleton opportunity. A cooperative agreement may still be
beneficial relative to an interacting noncooperative baseline. That
benchmark differs from removing all other travellers.

\subsection{4.1 Exact budget repair and approximate
stability}\label{exact-budget-repair-and-approximate-stability}

Allow an external subsidy \(\sigma\ge0\), so participants pay
\(x(N)=v(N)-\sigma\) while the original coalition departure
opportunities remain available. This is an explicit financing extension,
not a hidden interpretation of binding agreements.

\textbf{Theorem 5 --- Minimum subsidy.} The minimum stabilizing subsidy
for a singleton-dominated cost game is \(G\). At \(\sigma=G\), the
allocation \(x_i=v(\{i\})\) is stable. Every stable subsidized
allocation requires \(\sigma\ge G\).

\emph{Proof.} The proposed allocation satisfies every coalition bound by
(6) and has the required subsidized budget. Conversely, summing the
singleton bounds gives \(v(N)-\sigma\le\sum_i v(\{i\})\). \(\square\)

Alternatively retain the full budget and permit a per-member relaxation:

\[x(N)=v(N),\qquad x(S)\le v(S)+|S|\varepsilon. \tag{9}\]

\textbf{Theorem 6 --- Optimal uniform relaxation.} For \(n=|N|>0\), the
minimum tolerance in (9) is \(G/n\). The allocation

\[x_i=v(\{i\})+G/n \tag{10}\]

attains it.

\emph{Proof.} Equation (10) is efficient. Summing it over \(S\) and
using (6) establishes (9). For any feasible allocation in (9), summing
the singleton inequalities gives \(G\le n\varepsilon\). \(\square\)

For the thesis table, \(G=1\): a subsidy of 1 stabilizes payments
\((4,4)\), or the full budget admits the optimal tolerance \(1/2\) with
allocation \((4.5,4.5)\). These are TU financing/allocation conclusions.
Neither allocation is silently asserted to be a deterministic attainable
NTU physical-cost vector.

The cost of stability and weighted/per-capita least-core approaches are
established concepts {[}5,6{]}. Our relaxation is explicitly nonnegative
and includes all coalitions; this avoids ambiguity with conventions that
optimize surplus only over proper nontrivial coalitions. The elementary
formulas above are verified consequences for the declared class, not
claims to invent those concepts.

\subsection{4.2 An admissible financed isolation
counterpart}\label{an-admissible-financed-isolation-counterpart}

There is also a conditional repair that retains isolated departure
opportunities. It changes financing explicitly rather than treating a
positive budget gap as removable by redistribution alone. A sponsor
commits a subsidy \(G\) to full-team service; departing proper
coalitions retain their original isolated costs and do not receive that
subsidy. Define the participant-budget game

\[v^G(S)=\begin{cases}v(N)-G,&S=N,\\v(S),&S\ne N.\end{cases}\]

This game accounts for participants' net financing; the physical
grand-coalition service cost remains \(v(N)\). The sponsor is external
to the player set, with no budget or participation guarantee supplied by
the theorem. For nonempty \(N\), the empty-coalition value is unchanged.

\textbf{Proposition 4.2.1B --- Financed isolation counterpart.} Consider
the monotone resource model (5), with attained coalition minima. Let
\(s\) be both a Nash equilibrium of the physical game and a physical
social optimum. Commit the fixed participation rebates

\[r_i=c_i(s)-v(\{i\}).\]

Then \(r_i\ge0\), \(\sum_i r_i=G\), the rebated strategic game
\(\widetilde c_i(t)=c_i(t)-r_i\) retains \(s\) as Nash, and its realized
net allocation at \(s\) belongs to the ordinary TU cost core of \(v^G\).

\emph{Proof.} Optimized solo cost is no greater than solo cost on the
same route, which by monotone congestion is no greater than its physical
cost. Thus \(v(\{i\})\le c_i(s)\). Physical optimality gives
\(C(s)=v(N)\), so the rebates sum to \(G\). Each rebate is fixed
independently of the eventual action profile, and cancels from both
sides of every unilateral Nash comparison. At \(s\), net costs equal
\(v(\{i\})\). Singleton domination gives every proper-coalition core
inequality, and \(\sum_i v(\{i\})=v(N)-G\) gives the funded grand
budget. These steps, including the resource-level theorem and conversion
from subsidized stability to the funded game's ordinary core, are
machine-verified. \(\square\)

The rebate is a fixed entitlement for participation within the specified
game; it is not recomputed after each deviation or promised after
leaving the mechanism. Full-team finance and isolated exit opportunities
must be implemented as specified. Truthful reporting and the sponsor's
incentive to pay are separate questions.

For Table 4.1, \(s=(p_1,p_2)\) has physical costs \((4,5)\), isolated
costs \((4,4)\) and \(G=1\). Rebates \((0,1)\) give net costs \((4,4)\)
and leave the split profile Nash. The funded values are
\(v^G(\{1\})=v^G(\{2\})=4\) and \(v^G(N)=8\), so the net allocation is
in its core. Physical service still costs \(v(N)=9\); participants cover
8 and the sponsor covers 1. The subsidy is minimal by Theorem 5. When
\(G=0\), the construction needs no financing and all rebates vanish,
recovering the unfinanced claim only under that restriction.

\section{5. Contract enforcement and its
limits}\label{contract-enforcement-and-its-limits}

The source assumes compliance with binding agreements in Chapter 3,
p.48, and permits the app to raise deviation prices in Chapter 6, p.133.
We make the required pricing capability explicit. For an announced
contract \(s\), define

\[p_i^s(t)=\max\{0,c_i(s)-c_i(t)\},\qquad
\widehat c_i^s(t)=c_i(t)+p_i^s(t). \tag{11}\]

The charge is nonnegative and is zero at the agreement. It requires
observation of the relevant profile costs and authority to impose the
charge, without a binding cap below the required amount. It is a
proposed completion of the referee mechanism, not a claim that the
original code computes this formula.

\textbf{Theorem 7 --- Implemented coalition stability.} Every contract
\(s\) is strong Nash under \(\widehat c^s\). More strongly, no coalition
can lower its aggregate implemented cost by a joint deviation. In a
finite game, the realized implemented costs therefore belong to a
fixed-outsider TU cost core whose opportunities remain subject to (11).

\emph{Proof.} For every \(i,t\), equation (11) implies
\(\widehat c_i^s(t)\ge c_i(s)=\widehat c_i^s(s)\). Consequently no
coalition can strictly improve all coordinates, or lower their sum. The
finite fixed-outsider minima exist, and Theorem 1 applies. \(\square\)

This is not dominant-strategy truthfulness, a credibility theorem for
the app, or stability after a coalition escapes its jurisdiction. Nor
does it show that \(s\) minimizes physical cost. In the original table,
the all-preferred contract costs 16 rather than 9, yet (11) strongly
implements it. Lean verifies both its inefficiency and its
enforceability. Choosing an efficient agreement is therefore a separate
task. If \(s\) is already a physical social optimum, (11) preserves its
on-path costs and implements it; the premise is stated, not inferred
from enforcement.

The mechanism also clarifies participation. A traveller may compare the
contract with an exit option outside the app. That outside opportunity
must enter a separate individual-rationality condition. The theorem does
not obtain consent from an absent or unspecified exit-payoff model.

\section{6. Repeated play, empirical equilibrium and observed
costs}\label{repeated-play-empirical-equilibrium-and-observed-costs}

For a fixed physical game, define the cumulative cost improvement of a
fixed alternative:

\[D_i(T,a_i)=\sum_{t=1}^T[c_i(s^t)-c_i(a_i,s_{-i}^t)]. \tag{12}\]

An external-regret bound \(R_i(T)\) satisfies \(D_i(T,a_i)\le R_i(T)\)
for \textbf{every} fixed action. It is an upper bound, not necessarily
the nonnegative clipping of regret. A calculation for only one preferred
alternative is insufficient.

The empirical distribution is

\[q_T(s)=\frac1T|\{t:s^t=s\}|,\qquad T>0. \tag{13}\]

We formally verify nonnegativity, normalization and the equality between
empirical expectation and uniform time average.

\textbf{Theorem 8 --- Empirical CCE certificate.} Let \(T>0\) and
suppose \(D_i(T,a_i)\le R_i(T)\) for every player and every fixed
action. Then

\[\mathbb E_{q_T}c_i(s)\le
\mathbb E_{q_T}c_i(a_i,s_{-i})+R_i(T)/T. \tag{14}\]

Thus \(q_T\) satisfies the CCE deviation inequalities with
player-specific bounds. Nonnegative \(R_i(T)/T\) are standard
approximate-CCE tolerances; a negative bound certifies a stronger
inequality. The distribution is an exact CCE when all fixed-action
deviation sums are nonpositive.

\emph{Proof.} Divide (12) by positive \(T\) and apply the empirical
expectation identity. The formal statement quantifies over all fixed
deviations. \(\square\)

A CCE compares deviations chosen independently of a recommendation. CE
permits recommendation-dependent deviation tests and is stronger.
Communication through an app does not supply either set of obedience
inequalities by itself. Equation (14) is a finite-time certificate. An
asymptotic topological convergence theorem is not part of the present
Lean library.

\subsection{6.1 Smoothness and physical
efficiency}\label{smoothness-and-physical-efficiency}

Assume the physical game is \((\lambda,\mu)\)-smooth:

\[\sum_i c_i(s_i^*,s_{-i})\le\lambda C(s^*)+\mu C(s)
\quad(\forall s,s^*),\qquad \lambda>0,\quad\mu<1. \tag{15}\]

\textbf{Theorem 9 --- Finite-time efficiency.} Suppose the game
satisfies (15), \(T>0\), and \(D_i(T,a_i)\le R_i(T)\) for every player
and every fixed action. For any common joint comparator \(s^*\),

\[\frac1T\sum_t C(s^t)\le
\frac{\lambda}{1-\mu}C(s^*)+
\frac{\sum_i R_i(T)}{T(1-\mu)}. \tag{16}\]

\emph{Proof.} Sum the regret inequalities for alternatives \(s_i^*\).
Apply (15) at each round to obtain

\[(1-\mu)\sum_t C(s^t)\le T\lambda C(s^*)+\sum_iR_i(T).\]

The denominator \(T(1-\mu)\) is positive. Division proves (16). The
comparator need not be an optimum; using one requires its existence and
feasibility. \(\square\)

This is the established smoothness/regret argument of Roughgarden
{[}4{]}, independently formalized here. It reconstructs the structure
suggested by thesis equations (6.5)--(6.13), pp.132--133 (PDF
pp.152--153). We use individual costs in (12), preserve one residual
sign and retain time sums. We do not label the original overloaded
expressions or the displayed equation (6.11) verified unchanged.

\subsection{6.2 Changing referee prices and
personalization}\label{changing-referee-prices-and-personalization}

The thesis allows personalized/system costs to differ and permits prices
to change across rounds. Let \(L_i^t(s)\) be the observed cost actually
optimized. It may include prices or subjective doctrine scores; physical
cost remains \(c_i(s)\). Set \(d_i^t(s)=L_i^t(s)-c_i(s)\). Suppose
observed external regret is at most \(\widehat R_i(T)\), and

\[\sum_t[d_i^t(a_i,s_{-i}^t)-d_i^t(s^t)]\le B_i(T)
\quad(\forall i,a_i). \tag{17}\]

\textbf{Theorem 10 --- Price-distortion bridge.} Physical regret is
bounded by \(\widehat R_i(T)+B_i(T)\). Therefore (16) holds with the
remainder

\[\frac{\sum_i[\widehat R_i(T)+B_i(T)]}{T(1-\mu)}. \tag{18}\]

A sufficient certificate for (17) is \(|L_i^t(s)-c_i(s)|\le\delta_i^t\)
for every profile, yielding \(B_i(T)=2\sum_t\delta_i^t\).

\emph{Proof.} For each fixed alternative,

\[c_i(s^t)-c_i(a_i,s_{-i}^t)=L_i^t(s^t)-L_i^t(a_i,s_{-i}^t)
+d_i^t(a_i,s_{-i}^t)-d_i^t(s^t).\]

Sum this identity and apply both bounds. The absolute-error certificate
bounds each distortion difference by \(2\delta_i^t\). Apply Theorem 9 to
the resulting physical regret bound. \(\square\)

An action-independent fee cancels in (17). A bounded but persistent
action-dependent distortion can instead produce a linear \(B_i(T)\),
leaving a nonvanishing efficiency remainder. Contract enforcement
illustrates the issue concretely: at the all-preferred table profile,
each traveller has a physical unilateral improvement of 3, while (11)
makes the implemented improvement zero. These two numerical statements
are machine-verified. Observed obedience or observed no-regret is
therefore not interchangeable with physical no-regret.

Theorem 10 allows arbitrary round-dependent observed costs, but retains
a fixed player set, action sets and physical base game. Participant
turnover or a time-varying physical environment requires an additional
comparator model and is not covered by changing notation.

\subsection{6.3 Social cost and pure price of
anarchy}\label{social-cost-and-pure-price-of-anarchy}

Social cost is the declared sum \(C(s)\), whereas pure price of anarchy
compares the worst pure Nash equilibrium with an unrestricted social
optimum. For \(C^*=\min_s C(s)>0\) and a nonempty pure-equilibrium set,

\[\mathrm{PoA}_{\mathrm{pure}}=\max_{s\in\mathrm{PNE}}\frac{C(s)}{C^*}. \tag{19}\]

Summing the unilateral Nash inequalities and applying (15) gives
\(C(s)\le\lambda C^*/(1-\mu)\) at every pure equilibrium. We verify this
direct corollary as \texttt{nash\_social\_efficiency}; it is the
zero-regret specialization of the standard argument. Its application
requires smoothness of the actual cost model. V's approved local
endpoint in Section 7 need not be Nash, so its ratio to \(C^*\) is an
optimization gap rather than a price-of-anarchy statistic.

We enumerated every pure profile and unilateral deviation in the 44
source-based cost tables and the 48 distinct broader-network games. The
latter are counted once per game, rather than twice for their two
baselines. All 92 games have pure equilibria. Among the 88 games with
positive optimum, the largest pure price of anarchy is 5: the
four-traveller layered instance with seed 2018101 has optimum 2 and
worst-equilibrium cost 10. Within the 44 source-based reconstructions,
the largest ratio is \(232/221\); the unperturbed coalesced vertex and
edge cases have ratios 1 and \(75/74\). These are exact finite
computations under the declared costs, not uniform bounds for a model
class.

Four broader games have zero optimal cost. In the ring-with-shortcuts
instance with four travellers and seed 2018101, profile 9 is Nash with
costs \((3,0,0,0)\), while profile 10 has costs \((0,0,0,0)\).
Consequently, no finite \(K\) can satisfy \(C(s)\le K C^*\) at every
pure equilibrium of that game. We kernel-verify this directly from its
weighted resource routes and private penalties; the independent checker
also matches the literal resource encoding to the saved instance. Zero
here means no overlap or preference charge, not zero travel time. In the
other three zero-optimum games, every pure equilibrium has zero cost;
their \(0/0\) ratios are left undefined. V reaches profile 10 from both
recorded baselines of the counterinstance.

The obstruction persists without a zero denominator. As an explicit
model variation, add a fixed private cost \(\varepsilon>0\) to each
action of each of the four travellers. Every cost remains nonnegative,
unilateral incentives are unchanged, profile 10 is still socially
optimal with cost \(4\varepsilon\), and profile 9 remains Nash with cost
\(3+4\varepsilon\). Its equilibrium cost ratio is

\[1+\frac{3}{4\varepsilon},\]

which is unbounded as \(\varepsilon\) decreases. Lean verifies the
stronger quantified statement: for every \(K\ge0\), a positive
\(\varepsilon\) yields a nonnegative-cost game with positive optimum and
a Nash equilibrium costing more than \(K\) times that optimum. Thus
monotone overlap costs and private preferences alone provide no uniform
finite pure-price-of-anarchy bound in this declared resource class. The
positive-base-cost family is an extension of the stored instance, not an
alteration of its recorded experimental costs.

Lean also verifies that the unshifted counterinstance cannot satisfy
(15) for any real \(\lambda\) and \(\mu<1\). Theorem 9 remains a valid
conditional guarantee; this example identifies a model for which its
hypothesis fails. The original territory-sharing formulation requires
its own smoothness and regret certificates. The selected
counterinstance, shifted-family proof and exhaustive equilibrium results
are retained in \texttt{ThesisLean/EfficiencyCase.lean} and
\texttt{research/efficiency/}.

\section{7. A bounded engineering guarantee and certification
interface}\label{a-bounded-engineering-guarantee-and-certification-interface}

\subsection{7.1 Participation-safe finite
selection}\label{participation-safe-finite-selection}

The engineering objective does not require that every negotiated outcome
belong to every possible cooperative core. A more direct guarantee is
that coordination preserves each participant's declared acceptance bound
and optimizes joint cost within those bounds. Let \(b\) be a feasible
interacting baseline, and let \(h_i\) be traveller \(i\)'s accepted cost
ceiling. Require \(c_i(b)\le h_i\), and define

\[F_h=\{s\in A:c_i(s)\le h_i\text{ for every }i\}.\]

\textbf{Theorem 11 --- Participation-safe selection.} In a finite game,
\(F_h\) is nonempty and contains a minimum-total-cost profile \(s_h\),
with \(C(s_h)\le C(b)\). This joint non-worsening holds for every
ceiling vector admitting \(b\). In particular, setting \(h_i=c_i(b)\)
gives

\[c_i(s_h)\le c_i(b)\quad(\forall i),\qquad C(s_h)\le C(b),\]

and no other profile respecting those same bounds has a smaller joint
cost.

\emph{Proof.} The baseline witnesses nonemptiness. A real-valued
function on a nonempty finite set attains a minimum. Apply this to \(C\)
on \(F_h\). Membership gives the individual bounds, and comparison with
\(b\) gives joint non-worsening. Both the general ceiling result and its
baseline-safe specialization are machine-verified. \(\square\)

This theorem resolves existence of an acceptable coordinated plan for
this explicitly bounded engineering task. It does not promise a strict
improvement: returning the baseline may be necessary. Nor does it imply
unconstrained social optimality, strategic truthfulness or coalition
stability. The baseline includes other travellers' traffic; it is not
the inconsistent promise that everyone simultaneously obtains an
isolated solo optimum.

Assumption 4.3.1 in the thesis permits a traveller to volunteer a
less-preferred choice {[}1, p.56, PDF p.76{]}. A declared nonnegative
sacrifice budget \(\eta_i\) can be represented by \(h_i=c_i(b)+\eta_i\).
An individual may then incur a higher cost within the accepted
allowance. Exact minimum-total-cost selection still guarantees
\(C(s_h)\le C(b)\), because \(b\) remains an approved candidate.
Enlarging the approved set can only weakly lower its minimum joint cost.
Approval alone does not give this joint guarantee for an arbitrary
heuristic proposal; such a procedure needs its own objective acceptance
rule. Theorem 12 supplies that rule through strict descent. Consent is a
specification input, not something inferred from a threat score.

A concrete conservative completion is to enumerate feasible profiles,
reject those exceeding a declared ceiling, and select a survivor of
minimum joint cost. Exhaustive finite enumeration terminates, but its
search size is \(\prod_i|A_i|\) and no scalability claim follows. A
reference implementation is supplied separately; it is not a formally
verified extraction of the Lean proof. On all four possible baselines of
Table 4.1, it selects a split profile with total cost 9 and does not
increase either traveller's cost. A bargaining heuristic can propose
candidates to this selection layer, but without exhaustive search or a
separate optimality certificate its guarantee is only acceptance of a
checked proposal and fallback to the baseline. This completion does not
assert termination or optimality of the original Algorithms 1--2.

\subsection{7.2 Reusing the original negotiation
results}\label{reusing-the-original-negotiation-results}

The original ten-traveller study supplies a concrete participation
example {[}1, Table 4.4, p.70, PDF p.90{]}. Its displayed total costs
progress from 320 to 308, 300 and finally 282. The final cost vector is
coordinatewise no greater than the initial vector, with a total
reduction of 38 (11.875\%). Some provisional rounds exceed individual
baseline costs: travellers 8 and 10 in the first round, and 2, 8 and 10
in the second. The example therefore motivates a distinction between
provisional negotiation and final acceptance. It illustrates the
baseline bounds of Theorem 11, without asserting that the displayed
final agreement is the minimum over all approved plans.

The thesis already characterizes the method as heuristic and potentially
nonoptimal {[}1, §4.5.4, pp.74--76, PDF pp.94--96{]}. In its Table 4.7,
all 20 reported bargaining outcomes improve on their selfish baselines;
two equal the reported optimum. The displayed bargaining-to-optimum
ratios range from 1 to 1.088, with one to four negotiation rounds. We
checked these arithmetic statements against transcribed table entries
and visually checked the source tables. These are attributed historical
observations, not rerun experiments, independently certified optima, or
worst-case approximation and runtime guarantees.

The reusable algorithmic structure is the strategy broadcast,
conflict-based partner selection, bilateral proposals, joint-cost
comparison and coalition updates in Algorithms 1--2 {[}1, pp.55--57, PDF
pp.75--77{]}. A precise implementation can use that proposal mechanism
with a participant acceptance filter and a strict joint-cost descent
rule for committed updates. Section 7.3 supplies finite sequential
scheduling and explicit tie/no-change rules, together with a termination
theorem for that completion. The original permission to reconsider
harmful coalitions {[}1, pp.60--61 and 68--71{]} supports such checks,
while the intermediate cost increases show why a claim of individual
improvement at every provisional step would misrepresent the procedure.

\subsection{7.3 An executable finite negotiation
procedure}\label{an-executable-finite-negotiation-procedure}

We now specify a conservative completion of the bilateral proposal
mechanism. Its inputs are an ordered finite list \(F\) of feasible
complete plans, a rational-valued joint-cost oracle \(J\), an approval
predicate \(H\) and a permitted-move predicate \(M\). A checked entry
point rejects an infeasible or unapproved baseline. From the current
committed plan \(s\), scan \(F\) and accept the first \(t\) satisfying

\[H(t),\qquad M(s,t),\qquad J(t)<J(s).\]

Commit \(t\) and restart a full scan. If no candidate passes, stop. List
order resolves proposal priority and equal-cost proposals are rejected.
Restarting the scan matters: a proposal previously rejected by \(M\) can
become permitted after another commitment. In the binary-route
implementation, \(M\) permits one or two travellers to change actions
and \(H\) checks every traveller's baseline ceiling. These rules reuse
the proposal, joint-cost comparison and acceptance structure of
Algorithms 1--2. They explicitly fix sequential scheduling and strict
commitment rules; the implementation does not purport to reconstruct the
original numerical threat policy or coalition-representative election.

The executable control flow can be summarized as follows:

\begin{Shaded}
\begin{Highlighting}[]
\NormalTok{check that baseline b is listed and approved}
\NormalTok{s := b}
\NormalTok{for at most |set(F)| iterations:}
\NormalTok{    scan F for the first approved, permitted t with J(t) \textless{} J(s)}
\NormalTok{    if no such t exists: return s}
\NormalTok{    s := t}
\NormalTok{return s}
\end{Highlighting}
\end{Shaded}

The final return is safe: the following theorem proves that sufficient
finite-state fuel cannot leave an eligible improvement unchecked at the
returned state.

\textbf{Theorem 12 --- Finite negotiation correctness.} Let the baseline
belong to \(F\) and satisfy \(H\). The procedure terminates with an
approved feasible plan \(s_f\), with \(J(s_f)\le J(s_0)\). At
termination,

\[H(t)\ \wedge\ M(s_f,t)\quad\Longrightarrow\quad J(s_f)\le J(t)
\qquad(t\in F).\]

A bound of \(|\operatorname{set}(F)|\) iteration positions suffices,
including a terminal certificate. The statement does not require
distinct costs, integer costs or a positive lower bound on costs.

\emph{Proof.} Define the rank of \(s\) to be the number of distinct
listed plans with objective strictly less than \(J(s)\). For every
accepted transition \(s\to t\), the profiles cheaper than \(t\) form a
proper subset of those cheaper than \(s\): inclusion follows from strict
descent, and \(t\) belongs only to the second set. Rank therefore
strictly decreases. Induction on the supplied finite-state bound proves
that the executable result admits no eligible successor; the bound
cannot expire at a nonterminal state. A second induction proves
feasibility, approval and joint-cost non-worsening. The finite scan's
correctness equates absence of an eligible successor with the displayed
terminal inequalities. All these claims are proved for the executable
Lean definitions. A semantic bridge transfers approval and cost
guarantees to the earlier real-valued planning model when the rational
oracle and approval test accurately represent that model. \(\square\)

The Lean implementation executes directly, and a Lake executable uses
the same definitions. Kernel computations verify all four starting
profiles of Table 4.1; the cost-oracle encoding is also proved equal to
the original formal table. Starting from either diagonal profile it
selects a split profile of total cost 9, and a split baseline is
retained. A further source-specific theorem proves that every returned
plan in these four runs is physically Nash and jointly optimal,
independently supplying the stronger premises used by the cooperative
counterparts. Combining those premises with the earlier fixed-outsider
theorem proves that the actual returned plan admits the corresponding TU
core allocation. An independently written exact-arithmetic Python
reference checks every baseline of all 6,561 two-player cost tables with
individual costs in \(\{0,1,2\}\), totaling 26,244 runs. This
operational check supplements the general Lean proof; the Python program
is not claimed to be verified extraction.

The endpoint guarantee is deliberately local to approved permitted
moves. Two separate synthetic examples are kernel-checked: a
three-person change can reach a cheaper approved plan that no pairwise
move reaches, and a terminal agreement can admit an individual physical
improvement that violates another traveller's bound. Consequently
Theorem 12 is not a global-optimality or Nash theorem. Stronger
cooperative conclusions still require the hypotheses of Propositions
4.2.1A--B or an independently checked coalition certificate. The current
complete scan can be exponential in the number of travellers; no
scalability result is asserted.

\subsection{7.4 Reconstructing the ten-traveller heuristic on common
instances}\label{reconstructing-the-ten-traveller-heuristic-on-common-instances}

The original raw data and implementation are unavailable. We therefore
implement an explicitly documented \emph{reconstruction of the thesis
heuristic} and compare it with the executable verified procedure on
identical reconstructed cost tables. This is a new computational study,
not a replication of the reported negotiation trace.

We transcribe the twenty routes in (4.24)--(4.43) {[}1, pp.65--66, PDF
pp.85--86{]}. The primary model treats the primed endpoint labels
printed for traveller 6 as aliases of the corresponding unprimed
endpoints. Literal-prime sensitivity cases retain them as distinct
resources. For each of two resource conventions---route vertices and
directed route edges---we use

\[c_i(s)=p_i a_i+\sum_{e\in U_i(a_i)}w_e\bigl(n_e(s)-1\bigr),\]

where \(a_i\in\{0,1\}\) and every used resource has load at least one.
The unperturbed cases use \(w_e=1\) and \(p_i=0\), consistent with
sponsored route base costs and participant overlap charges. Unit vertex
overlaps with coalesced endpoints reproduce all twenty components of
Table 4.2's all-first and all-second cost columns without estimated
numeric weights. Edge overlaps do not. This endpoint agreement is a
calibration check, not independent identification of the original model.
The later reported negotiation vectors are not recovered by these simple
conventions, so we retain both resource choices and disclose the label
normalization rather than infer an undocumented original cost oracle.

The primary heuristic H1 selects conflicting coalition partners by
shared-resource count, elects a representative by unilateral cost-change
risk, and compares offers for either representative to switch routes.
The offer score uses the minimum of the two representatives' absolute
cost changes, following the structure of (4.1); lower scores are
preferred, with the initiator's offer retained on a tie. A provisional
commitment requires improved combined negotiating-coalition cost and
non-increased total cost. Coalitions merge on agreement; individually
profitable withdrawals are reconsidered. A visited-profile safeguard
prevents cycles and is an explicit reconstruction choice. Participant
baseline bounds are checked at the end, with baseline fallback available
if approval fails. These rules do not claim a uniquely recovered threat
policy or mixed-retaliation distribution. H2 changes the risk score to
the switching representative's own change; H3 reverses scheduling and
partner-priority conventions.

For each instance, we enumerate all \(2^{10}=1,024\) profiles to compute
both the global joint-cost minimum and the minimum among
baseline-approved profiles. The verified comparator V uses the actual
Lean executable on the same integer cost table; its results agree with
the instrumented Python reference. An independent checker recomputes
costs through pairwise resource intersections and checks approval, all
unilateral deviations, every approved one/two-person improvement and the
reported exact minima.

On the coalesced vertex instance, the all-first baseline costs 320. H1
and V both attain the computed global and approved minimum 254, through
four and six commitments respectively; they select different optimal
assignments. On the coalesced edge instance, the baseline is 178 and
both reach the computed minimum 148 in five commitments. The
literal-prime baselines and minima are respectively \(302\to242\) for
vertices and \(174\to144\) for edges. All four unperturbed cases finish
with participant approval and physically Nash assignments under their
declared models. These numbers are reconstructed results; 254 is not
presented as the original experiment's recovered optimum.

We also retain all twenty fixed-seed perturbations per coalesced
resource model, with positive integer resource weights 1--3 and
alternative-route preference penalties 0--3. The resulting 42
primary/perturbed cases and two literal-prime cases are a small
controlled family, not a representative sample of transport networks.
The following counts concern the 21 coalesced cases in each resource
model:

\begin{longtable}[]{@{}
  >{\raggedright\arraybackslash}p{(\columnwidth - 10\tabcolsep) * \real{0.1364}}
  >{\raggedright\arraybackslash}p{(\columnwidth - 10\tabcolsep) * \real{0.1364}}
  >{\raggedleft\arraybackslash}p{(\columnwidth - 10\tabcolsep) * \real{0.1818}}
  >{\raggedleft\arraybackslash}p{(\columnwidth - 10\tabcolsep) * \real{0.1818}}
  >{\raggedleft\arraybackslash}p{(\columnwidth - 10\tabcolsep) * \real{0.1818}}
  >{\raggedleft\arraybackslash}p{(\columnwidth - 10\tabcolsep) * \real{0.1818}}@{}}
\toprule\noalign{}
\begin{minipage}[b]{\linewidth}\raggedright
Resource model
\end{minipage} & \begin{minipage}[b]{\linewidth}\raggedright
Procedure
\end{minipage} & \begin{minipage}[b]{\linewidth}\raggedleft
Approved
\end{minipage} & \begin{minipage}[b]{\linewidth}\raggedleft
Physical Nash
\end{minipage} & \begin{minipage}[b]{\linewidth}\raggedleft
Approved pairwise terminal
\end{minipage} & \begin{minipage}[b]{\linewidth}\raggedleft
Computed global minimum
\end{minipage} \\
\midrule\noalign{}
\endhead
\bottomrule\noalign{}
\endlastfoot
Vertex & H1 & 21/21 & 21/21 & 16/21 & 16/21 \\
Vertex & H3 & 21/21 & 21/21 & 18/21 & 18/21 \\
Vertex & V & 21/21 & 21/21 & 21/21 & 21/21 \\
Edge & H1 & 21/21 & 19/21 & 11/21 & 11/21 \\
Edge & H3 & 21/21 & 21/21 & 17/21 & 17/21 \\
Edge & V & 21/21 & 21/21 & 21/21 & 21/21 \\
\end{longtable}

H2 coincides with H1 throughout this collection. Scheduling sensitivity
changes the heuristic outcomes, supporting the need to state
reconstruction choices. V's global optimality and physical Nash status
in these runs are observed finite-instance properties, not strengthened
conclusions of Theorem 12. All methods improve joint cost in the tested
cases, but the stronger terminal certificate distinguishes V from some
heuristic endpoints. We count commitments, not the original simulation's
communication rounds. Common exhaustive cost-table construction is
excluded from lookup-phase timing, and no distributed speedup or
scalability claim follows. All inputs, seeds, code, complete tables for
the four unperturbed cases, traces and independent certificate checks
accompany the paper. The larger-instance numerical properties are exact
computational evidence; they are not additional Lean-kernel theorems.

Figures 2 and 3 show complementary aspects of the reconstructed
comparisons. Figure 2 locates each commitment relative to the computed
cost minimum on the primary vertex instance; its horizontal axis counts
accepted updates rather than elapsed time. Figure 3 plots each
instance's percentage cost gap relative to its exact finite minimum. The
zero gaps for V are observations in this collection; Theorem 12
guarantees local terminal acceptance, not universal global optimality.

\begin{figure}
\centering
\includegraphics[width=6.3in,height=\textheight]{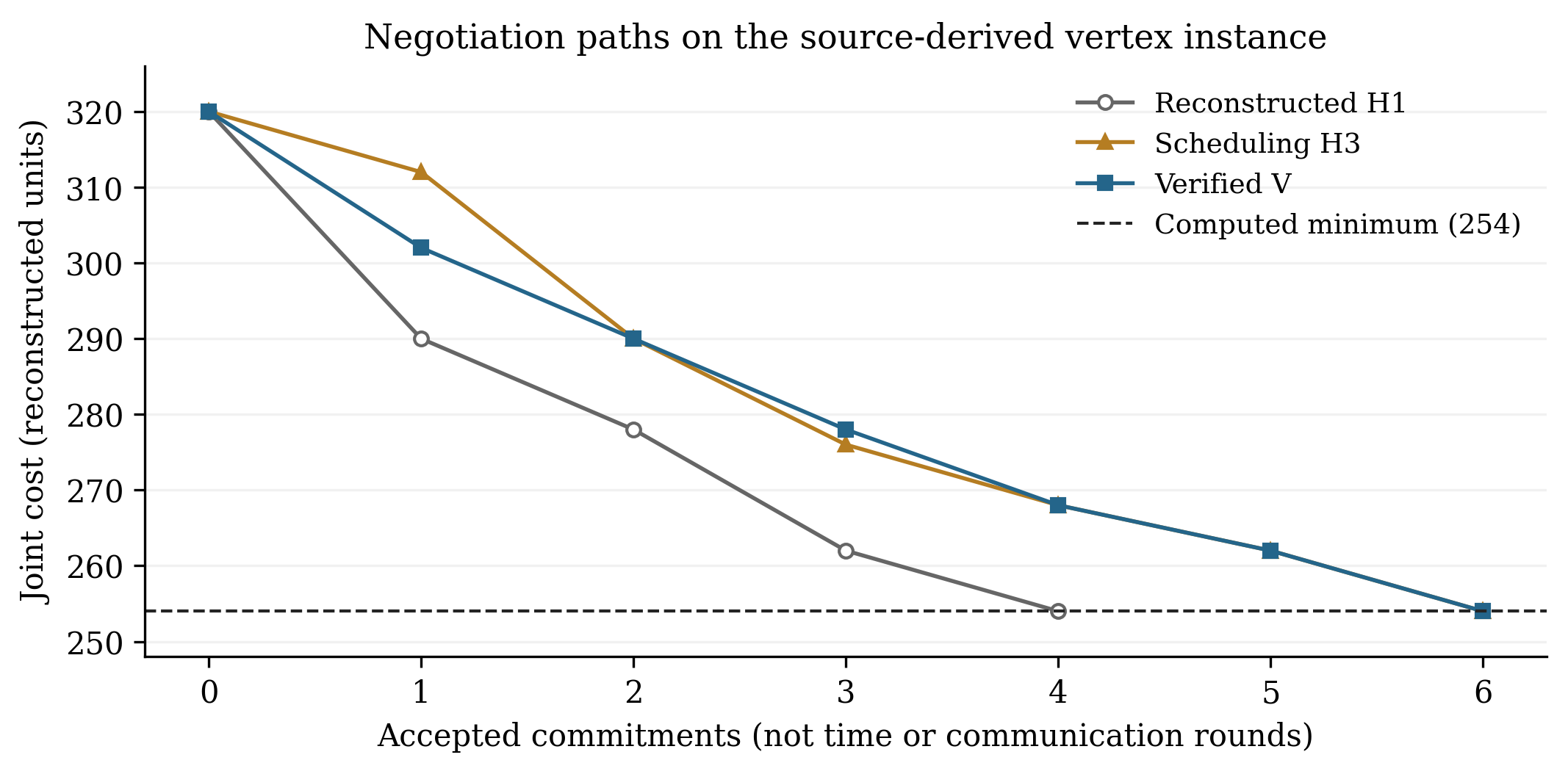}
\caption{Joint-cost paths on the source-derived vertex instance. The
paths are newly reconstructed executions, not the published negotiation
trace. All shown paths reach the computed minimum 254; accepted-update
counts are not communication rounds or runtime.}
\end{figure}

\begin{figure}
\centering
\includegraphics[width=6.3in,height=\textheight]{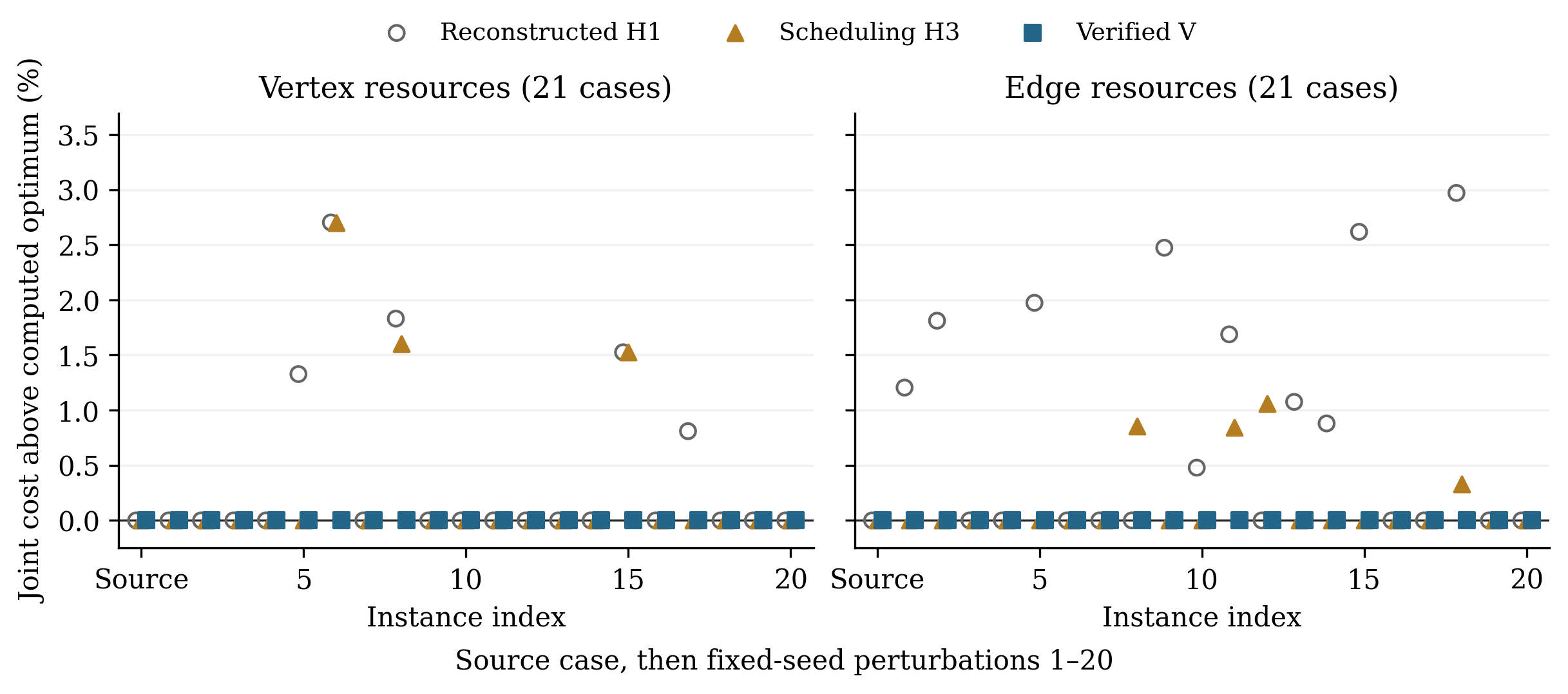}
\caption{Joint-cost gaps above the computed optimum for the 21 coalesced
cases per resource model. Markers are horizontally offset for
readability. Seed indices are discrete case identifiers; no statistical
confidence intervals or general approximation guarantee are implied.}
\end{figure}

\subsection{7.5 Independent network
families}\label{independent-network-families}

To test dependence on the original route skeleton, we fixed a generation
protocol before inspecting outcomes and retained all 48 generated
graph/route/cost instances. Four graph families cover a bidirectional
\(4\times4\) grid, a 12-node ring with diameter shortcuts, a directed
four-layer network and three parallel corridors. Each has 4, 6 or 8
travellers and four declared seeds. Two starting profiles per instance
(all-first and a seeded nonzero profile) give 96 baseline-specific
comparisons. These two observations share a cost table and are not
independent statistical samples.

Each traveller receives two distinct simple routes between its origin
and destination: a shortest path and a shortest alternative obtained by
deleting an edge of the first path. Directed edges are resources;
weights 1--3 and alternative-route private penalties 0--3 use the same
additive overlap-cost model as Section 7.4. Thus topology, route
overlap, traveller count and starting plan vary, while the cost class
and two-action restriction remain controlled. These newly generated
synthetic experiments supply a fresh test of the thesis-inspired
coordination logic on inputs independent of the original numerical
instances. Their value is the ability to vary network structure and
starting conditions while holding the declared cost class and comparison
rules fixed. They are a controlled robustness study rather than a
representative transport-network sample. The complete protocol, graphs,
routes, seeds and parameters are retained in
\texttt{research/broader\_networks/}.

All methods receive identical tables and baselines, but the original
comparison uses different acceptance policies: V applies participant
ceilings at every committed step, whereas H1 checks them only at the
endpoint and may fall back to the baseline. Consequently, the counts
compare complete procedures under those policies, rather than isolate
algorithmic superiority. The matched controls below address this
difference. Exhaustive enumeration separates the unrestricted minimum
\(C^*\) from the approved minimum \(C_A^*\). For a published approved
endpoint \(s\), the gap decomposes exactly as

\[C(s)-C^* = \underbrace{C_A^*-C^*}_{\text{participation gap}} + \underbrace{C(s)-C_A^*}_{\text{search gap}}.\]

The first term measures the cost of the declared individual acceptance
bounds; the second measures residual suboptimality within those bounds.
A raw heuristic outcome violating approval is replaced by the baseline
for publication, as specified in Section 7.4. Raw approval and raw Nash
flags remain reported separately; no fallback result is relabelled as an
achieved negotiated improvement.

\begin{longtable}[]{@{}
  >{\raggedright\arraybackslash}p{(\columnwidth - 10\tabcolsep) * \real{0.1304}}
  >{\raggedleft\arraybackslash}p{(\columnwidth - 10\tabcolsep) * \real{0.1739}}
  >{\raggedleft\arraybackslash}p{(\columnwidth - 10\tabcolsep) * \real{0.1739}}
  >{\raggedleft\arraybackslash}p{(\columnwidth - 10\tabcolsep) * \real{0.1739}}
  >{\raggedleft\arraybackslash}p{(\columnwidth - 10\tabcolsep) * \real{0.1739}}
  >{\raggedleft\arraybackslash}p{(\columnwidth - 10\tabcolsep) * \real{0.1739}}@{}}
\toprule\noalign{}
\begin{minipage}[b]{\linewidth}\raggedright
Family (24 starts each)
\end{minipage} & \begin{minipage}[b]{\linewidth}\raggedleft
H1 raw approved
\end{minipage} & \begin{minipage}[b]{\linewidth}\raggedleft
V approved
\end{minipage} & \begin{minipage}[b]{\linewidth}\raggedleft
V approved minimum
\end{minipage} & \begin{minipage}[b]{\linewidth}\raggedleft
V unrestricted minimum
\end{minipage} & \begin{minipage}[b]{\linewidth}\raggedleft
V underlying-cost Nash
\end{minipage} \\
\midrule\noalign{}
\endhead
\bottomrule\noalign{}
\endlastfoot
Grid & 19 & 24 & 23 & 18 & 22 \\
Ring with shortcuts & 12 & 24 & 23 & 13 & 17 \\
Layered & 11 & 24 & 22 & 12 & 14 \\
Parallel corridors & 14 & 24 & 24 & 14 & 14 \\
\end{longtable}

V preserves approval and reaches an approved pairwise terminal point in
all 96 cases, attaining the exact approved minimum in 92 and the
unrestricted minimum in 57. It is physically Nash in 67. Joint cost
strictly improves in 66 cases and the baseline is retained in 30; the
participation gap is positive in 38. H1's raw outcomes are approved in
56 cases and physically Nash in 80; with baseline fallback, 46 published
outcomes attain the approved minimum and 22 the unrestricted minimum. H2
and H3 attain respectively 44 and 46 approved minima after fallback. The
higher raw Nash count of H1 concerns a different property and includes
outcomes some participants would reject. We do not claim that V
dominates every criterion or reproduces the original heuristic.

\begin{figure}
\centering
\includegraphics[width=1\textwidth,height=\textheight]{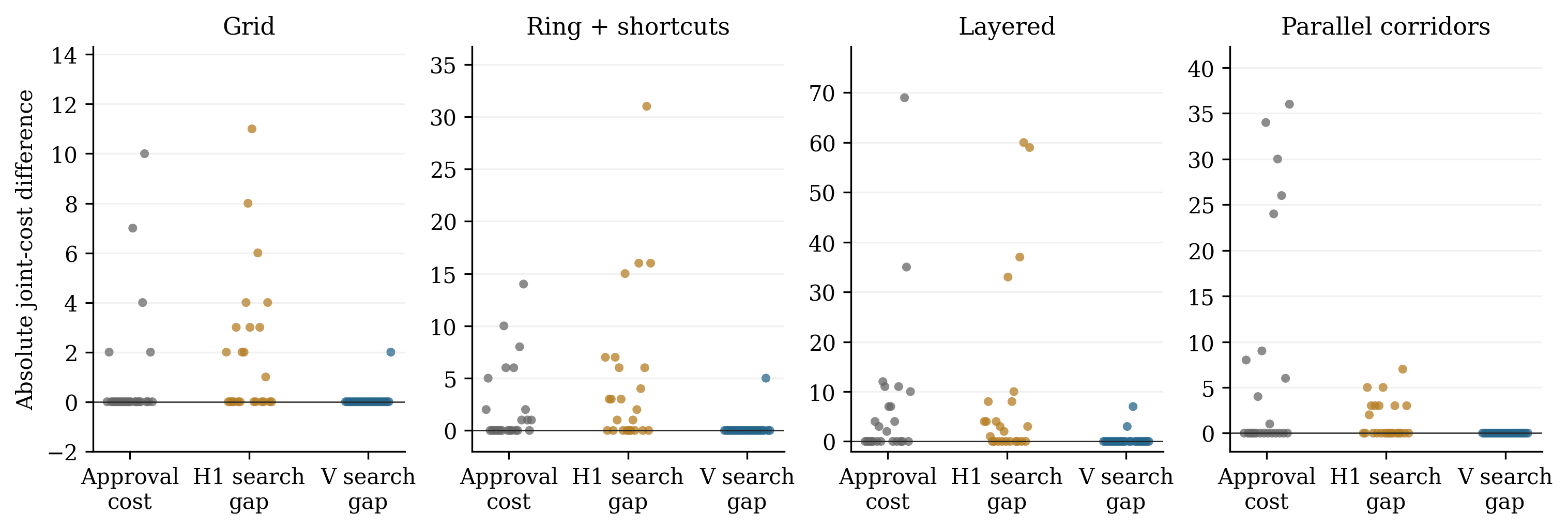}
\caption{Gap decomposition on independent network families. Each group
contains 24 baseline-specific cases. Approval cost is the participation
gap; H1 and V search gaps compare published outcomes with the exact
approved minimum. H1 includes baseline fallback when raw approval fails;
matched controls are reported separately below. Vertical scales differ
between families; horizontal offsets only reveal repeated values.}
\end{figure}

The four non-optimal V endpoints have search gaps 2, 5, 3 and 7 in the
declared planning-cost units. The largest relative search gap is 50\%,
so the high optimum count does not establish a uniform approximation
guarantee. One explicit grid case has eight travellers, seed 2018103 and
starting profile 214: V stops at profile 2 with cost 16, whereas
approved profile 216 costs 14. They differ in five traveller actions,
beyond the permitted pairwise move. This is a local-search limitation in
the reconstructed model, not a counterexample to Theorem 12.

A separate six-traveller grid case (seed 2018102, starting profile 49)
returns profile 33 with costs \((8,0,0,0,0,5)\), which is a minimum-cost
approved plan of cost 13. Traveller 1 can individually switch to profile
32, reducing its cost to 3 while raising traveller 4's cost from 0 to 3.
The latter exceeds that traveller's baseline ceiling of zero. Thus
approved optimality does not imply individual Nash stability: the
referee's permitted joint commitments and an individual's unenforced
deviation opportunities differ. The retained counterinstances expose
this distinction concretely, rather than concealing it behind favorable
aggregate results.

An independent checker recomputes all 39,936 player/profile costs by
pairwise intersections, validates simple routes and graph edges, and
checks 384 method endpoints and traces against exhaustive deviation and
minimum tests. The native Lean executable agrees with the reference in
all 96 cases. These are computational certificates under the declared
inputs; the unchanged generic protocol theorem remains kernel-verified.
The broader results add reproducible evidence for the engineering value
of coordinating personalized route choices with explicit participant
safeguards. They show how the thesis-inspired approach can be carried
into new instances through the revised mathematical treatment, with both
successful coordination and its limitations exposed by exact benchmarks.

\subsubsection{Retrospective matched
controls}\label{retrospective-matched-controls}

Following external feedback, we added acceptance-matched heuristic and
generic local-search controls on the same 96 baseline cases. The
protocol was fixed before running the new controls, but the instances
and old outcomes were already known; this is a retrospective diagnostic,
not a fresh validation sample. H1-A adds V's baseline-ceiling filter to
every H1 negotiation and withdrawal. H1-A-strict also requires strict
total-cost descent. All other representative, proposal and coalition
rules remain unchanged. Their proposal neighborhoods still differ from
V's full one-/two-person neighborhood.

\Needspace{8\baselineskip}

\begin{longtable}[]{@{}lrrrr@{}}
\toprule\noalign{}
Procedure & Raw approved & Raw Nash & Approved optimum & Mean gap \\
\midrule\noalign{}
\endhead
\bottomrule\noalign{}
\endlastfoot
H1 & 56/96 & 80/96 & 46/96 & 4.5833 \\
H1-A & 96/96 & 53/96 & 54/96 & 3.3958 \\
H1-A-strict & 96/96 & 52/96 & 54/96 & 3.4583 \\
V & 96/96 & 67/96 & 92/96 & 0.1771 \\
\end{longtable}

Only H1 requires endpoint fallback. Adding approval changes the
admissible trajectory and can reduce physical Nash incidence; the
objectives must therefore be distinguished even when all methods share
the filter. V attains more approved minima than these reconstructed
proposal mechanisms in this collection, but H1-A also beats V on one
instance: the eight-traveller grid case with seed 2018103 and baseline
214, returning cost 14 against V's 16. The outcomes therefore do not
establish pointwise dominance, identify a unique cause, or give a
general advantage.

A generic randomized pairwise local search uses exactly V's fixed
acceptance ceilings, objective and one-/two-person move set, choosing
uniformly among approved strictly improving neighbors. Across 256
retained schedules per case, its mean approved-optimum hit fraction from
the same baseline is \textbf{93.148\%}, compared with V's
\textbf{95.833\%} (92/96); mean absolute search gaps are respectively
0.3733 and 0.1771. A separate landscape diagnostic, starting uniformly
from approved profiles while keeping the original ceilings fixed, has a
mean hit fraction of \textbf{94.763\%} and mean gap 0.2535. Random
approved initialization may cross a multi-person barrier, so it is not
an operational substitute for the same-start control. Both randomized
arms attain the approved minimum on every trial in 85/96 cases. No
best-of-256 result or independent-trial significance claim is used.

The approved set has median size four, and in 20 cases it contains only
the baseline. The baseline already attains the approved minimum in 30
cases. These facts and the generic search controls show why 92/96 should
be read as finite implementation evidence rather than a distinctive
optimization achievement or strong evidence of general transfer. The
proved contribution of V is its explicit acceptance and stopping
guarantee. The synthetic study tests how this specification behaves in
the declared model class.

All 49,152 randomized traces and 384 deterministic endpoints pass
independent direct-cost checks, including approval, descent and
neighborhood conditions; randomized endpoints also pass exhaustive
local-terminal checks. The original H1/V outcomes are unchanged. Full
protocols, traces and certificates are retained in
\texttt{research/review\_controls/}.

\subsection{7.6 Certification
obligations}\label{certification-obligations}

The results support a mathematically specified certification layer
around personalized planning and negotiation. They do not require that
every bargaining process be declared convergent.

\textbf{Stage A: construct the game.} Record finite feasible actions,
full-profile physical costs, private observed costs and the common
social objective. State which hard constraints hold in every action; any
lottery or constraint relaxation must be explicit. For the resource
subclass, provide nondecreasing resource-price functions and private
route costs.

\textbf{Stage B: choose the opportunity model.} Specify fixed outsiders,
isolation or another response rule. For a fixed-outsider two-person
subgame, certify Nash and Pareto efficiency for an NTU claim, or Nash
and minimum joint cost for a TU claim. For larger coalitions, test the
actual coalitional condition rather than relying on pairwise agreement.
For isolation in the monotone subclass, compute the gap and use Theorems
4--6 to decide the required budget or tolerance.

\textbf{Stage C: separate selection from enforcement.} Negotiation
proposes an agreement. A referee may implement it through a declared
pricing rule such as (11). This certifies implemented stability under
the stated jurisdiction, not physical optimality, voluntary
participation or stability after exit. Quality claims must be checked
against physical costs and the appropriate outside options.

\textbf{Stage D: certify repeated outcomes.} Maintain the realized
history and evaluate every fixed action against actual opponents.
Compute upper bounds on observed regret and price distortion. Equations
(14) and (18) then certify empirical equilibrium and physical efficiency
when the recorded bounds are small. Smoothness must be proved for the
actual physical cost class, or checked exhaustively for a finite
instance. No smoothness certificate is inferred from the presence of a
referee.

This interface is operationally testable and does not encode the desired
outcome as an assumption. A regret certificate may be large; a
bargaining run may fail to meet a stopping criterion. Such failures
cannot be hidden by calling every final agreement an equilibrium.
Theorem 12 establishes termination for the specified finite completion.
Sublinear regret and guarantees for asynchronous or threat-based
variants require further algorithmic analysis. Standard
regret-minimizing learners offer a separate implementation route, with
their own information and bounded-loss assumptions {[}4{]}; their
guarantees are not attributed to Algorithms 1--2 without proof.

\section{8. Verification, prior art and research
significance}\label{verification-prior-art-and-research-significance}

The project kernel-checks the general theorems and declared
two-traveller instance results, including existence of finite coalition
minima and construction of the empirical distribution. The
larger-instance numerical results in Sections 7.4--7.5 use exact finite
computation with independent checks and are not additional kernel-proved
certificates. Every accepted module is imported by the root Lake target.
The verification script runs \texttt{lake\ build}, explicitly prints the
dependencies of every theorem and records source hashes. No admissions,
custom mathematical postulates, unsafe proof shortcuts or native
decision oracles are used. Reported foundational dependencies are
\texttt{propext}, \texttt{Classical.choice} and \texttt{Quot.sound}.
Noncomputable definitions over real numbers are mathematical
definitions, not executable learning algorithms.

Semantic verification is separate. The source table is reconstructed
from its chosen resource suffixes, and both cooperative conventions are
identified. The monotone model, penalty formula, subsidy and
observed-cost distortion assumptions are declared extensions. General
theorem validity does not prove that all fuzzy scores, rewards or
territory-sharing prices in the thesis instantiate those assumptions.
The original larger numerical experiments have not been replicated.
Section 7.4 instead supplies a controlled reconstruction with declared
cost, label and heuristic choices; its shared-instance comparisons are
computational evidence rather than a recovered original implementation.

The results draw on established theory. The finite-potential
minimization principle and congestion-game existence results are
established {[}2{]}. Fixed-outsider core correspondences express
established stability ideas under an explicit opportunity model.
Competitive equilibrium/core results for exchange economies, discussed
by Serrano {[}3{]}, do not automatically identify a route-choice Nash
equilibrium with a cooperative allocation. The financing and relaxation
results belong to the cost-of-stability and weighted least-core
frameworks {[}5,6{]}. Our particular formulas follow from the declared
singleton-dominance structure. The smoothness result is known {[}4{]};
the observed-cost extension follows by a perturbation identity. None is
presented as a new general solution concept.

The contribution is a formal mathematical treatment of the engineering
framework: personalized feasible plans can be coordinated within
explicit acceptance bounds, strategic agreements have cooperative
stability under the specified counterparts, contractual rules implement
the selected agreements, and repeated outcomes admit quantitative
efficiency certificates when their hypotheses hold. The original
two-traveller example instantiates both cooperative counterparts. The
purpose of this preprint is to present these definitions, proofs and
implementation obligations as a coherent foundation for coordinated
planning. The general mathematical results are attributed to their
established frameworks. The formal integration, executable procedure and
reproducible source-based comparisons are the contributions claimed
here; no new general equilibrium or bargaining solution concept is
asserted.

\subsection{Relationship to the 2018 engineering
framework}\label{relationship-to-the-2018-engineering-framework}

The new treatment strengthens the evidential basis of the 2018 thesis in
two complementary ways. First, it gives selected relationships between
individual choice, cooperative opportunities and enforcement explicit
mathematical conditions, together with machine-checked proofs. Second,
it turns thesis-inspired negotiation logic into inspectable
implementations and evaluates them on new synthetic data. The original
engineering idea---balancing personalized objectives with coordinated
use of shared resources---therefore receives both a precise formal
foundation and additional finite computational support.

The new data are useful precisely because they extend the investigation
beyond the original worked example. The 44 source-based reconstructions
test sensitivity to resource conventions, endpoint labels and costs. The
additional 48 graph/route/cost instances, evaluated from 96 starting
profiles across four network families, test whether the specified
coordination procedure continues to function when topology, overlap,
traveller count and starting conditions change. The generation protocol
was fixed before examining these outcomes, and every generated
configuration is retained. These experiments demonstrate execution
across the declared finite overlap-cost instances; their generators and
complete parameters make the evidence reproducible. The matched controls
show that the high optimum-hit count is also largely consistent with
generic pairwise local search, so it does not identify a distinctive
optimization benefit of the coordination mechanism.

The findings support a concrete engineering conclusion. Under the
revised procedure, coordination preserves every participant's baseline
cost ceiling and reaches an approved local terminal outcome in all 96
broader-network cases; 66 cases strictly improve joint cost and 30
retain the baseline. The exact approved minimum is attained in 92 cases,
with the local-search controls giving the necessary context for that
count. The reconstructed heuristic also produces joint improvements in
the source-based collection, while the broader comparisons reveal where
its provisional outcomes need an explicit acceptance safeguard.
Together, these results support the original motivation for combining
personal planning with negotiated coordination and show the practical
value of specifying the referee's obligations.

This is additional support for the original work through a new
mathematical and computational treatment. It does not retroactively
identify the revised procedure with every detail of the original
implementation. The new proofs establish the stated models and
procedure; the simulations demonstrate their behavior on the retained
instances. Four residual search gaps and 29 underlying-cost non-Nash
endpoints in the broader V runs delimit the stronger conclusions one can
draw. Thus the contribution is both positive and precise: the thesis's
engineering framework supports a rigorous, reproducible realization,
with its applicable guarantees and remaining limitations made explicit.

The study also illustrates the methodological opportunity that motivated
this return to the thesis. An engineer's intuition about competition and
cooperation can become the starting point for a fuller mathematical
investigation: reconstruct the intended model, make its assumptions
explicit, prove what follows, and test the resulting procedure on new
data. AI assistance makes that investigation accessible in this case;
Lean and reproducible experiments supply distinct checks on its results.
This is a documented case study of that process, with the mathematical
and engineering conclusions stated at the level supported by the
evidence.

The next substantive theorem should strengthen the finite negotiation
guarantee under realistic resource externalities, private information,
bounded referee prices and exit rights, or establish regret guarantees
for a specified learning rule. Stronger equilibrium claims cannot be
obtained by silently assuming these obligations away.

\textbf{Status of Proposition 4.2.1.} The source proposition does not
uniquely determine the required outsider opportunities. This addendum
proves two explicit conditional replacements. Proposition 4.2.1A proves
cooperative stability of efficient two-person Nash agreements under
fixed-outsider opportunities. Proposition 4.2.1B proves cooperative
stability under isolated opportunities with the exact external
financing, while preserving unilateral Nash incentives. Both are
realized by the original two-traveller example. Their Lean statements
are \textbf{VERIFIED}; their application to the engineering framework is
\textbf{CONDITIONALLY VERIFIED} under the assumptions stated in each
proposition. Together with participation-safe selection, these results
establish a rigorous coordination guarantee for the stated models.
Theorem 12 adds a terminating executable negotiation completion with an
approved local outcome. The reconstructed comparisons and new synthetic
experiments in Sections 7.4--7.5 add reproducible engineering evidence
on shared instances and across different network families. General
guarantees for heuristic variants, changing participant populations and
replication of the original empirical instances require separate
algorithmic or experimental evidence.

\subsection{Author contributions and AI
assistance}\label{author-contributions-and-ai-assistance}

Haitham Masaud Amar developed the original planning framework and
directed the present investigation.

OpenAI's Codex {[}11{]} assisted with mathematical reconstruction, Lean
proof development, experimental software, literature searches, figure
preparation, and manuscript drafting and revision. The author takes full
responsibility for the manuscript, including its mathematical claims,
formalization choices, references, and interpretation of the
computational results.

\subsection{Artifact availability and
reproduction}\label{artifact-availability-and-reproduction}

The companion \texttt{REPRODUCIBILITY\_BUNDLE.zip} contains the pinned
Lake manifest, all project Lean sources, executable entry points,
reconstruction inputs, plotting code and finite certificates. This draft
has no public archive identifier yet. With the pinned Lean toolchain
installed, the principal checks are:

\begin{Shaded}
\begin{Highlighting}[]
\ExtensionTok{lake}\NormalTok{ build ThesisLean negotiation reconstruction}
\ExtensionTok{lake}\NormalTok{ env lean ThesisLean/AxiomAudit.lean}
\ExtensionTok{python}\NormalTok{ scripts/reconstruct\_comparison.py }\AttributeTok{{-}{-}with{-}lean}
\ExtensionTok{python}\NormalTok{ scripts/check\_reconstruction.py}
\ExtensionTok{python}\NormalTok{ scripts/broader\_networks.py}
\ExtensionTok{python}\NormalTok{ scripts/check\_broader\_networks.py}
\ExtensionTok{python}\NormalTok{ scripts/review\_controls.py}
\ExtensionTok{python}\NormalTok{ scripts/check\_review\_controls.py}
\ExtensionTok{python}\NormalTok{ scripts/equilibrium\_efficiency.py}
\ExtensionTok{python}\NormalTok{ scripts/check\_equilibrium\_efficiency.py}
\end{Highlighting}
\end{Shaded}

The comparison scripts use the Python standard library. Regenerating
figures additionally uses Matplotlib and NumPy; manuscript exports use
Pandoc and pdfLaTeX. Source correspondence, complete assumption changes
and exact build records are retained in the companion research
directory.

\section{References}\label{references}

{[}1{]} H. M. Amar. \emph{Strategic Trip Planning: Striking a Balance
Between Competition and Cooperation}. PhD thesis, University of
Waterloo, 2018. Official PDF:
\href{https://dspacemainprd01.lib.uwaterloo.ca/server/api/core/bitstreams/8db00234-2b4d-4660-b7f2-d371de40a637/content}{Waterloo
repository}.

{[}2{]} R. W. Rosenthal. A class of games possessing pure-strategy Nash
equilibria. \emph{International Journal of Game Theory} 2, 65--67, 1973.
Source thesis reference {[}63{]}. The generic resource-potential
construction is identified as prior art; this development verifies the
general finite-potential implication and an explicit potential for Table
4.1.

{[}3{]} R. Serrano. Cooperative Games. In \emph{Computational
Complexity}, pp.666--674, 2012.
\href{https://doi.org/10.1007/978-1-4614-1800-9_42}{Publisher record}.
Related exposition: \emph{Cooperative Games: Core and Shapley Value},
CEMFI Working Paper 0709, 2007, §§2--3,
\href{https://www.cemfi.es/ftp/wp/0709.pdf}{working paper}.

{[}4{]} T. Roughgarden. Intrinsic robustness of the price of anarchy.
\emph{STOC}, 2009; extended \emph{Journal of the ACM} version, 2015.
\href{https://doi.org/10.1145/2806883}{DOI}.
\href{https://www.timroughgarden.algorithmsilluminated.org/papers/robust.pdf}{Author
version}.

{[}5{]} R. Meir, Y. Bachrach and J. S. Rosenschein. Minimal subsidies in
expense sharing games. \emph{SAGT}, 2010.
\href{https://www.microsoft.com/en-us/research/wp-content/uploads/2016/02/sagt10cos.pdf}{Author-hosted
paper}.

{[}6{]} T. Solymosi. Weighted nucleoli and dually essential coalitions.
\emph{International Journal of Game Theory} 48, 1087--1109, 2019.
\href{https://doi.org/10.1007/s00182-019-00689-x}{Open-access article}.

{[}7{]} J. F. Nash, Jr.~The Bargaining Problem. \emph{Econometrica}
18(2), 155--162, 1950. \href{https://doi.org/10.2307/1907266}{DOI}. The
product-maximization interpretation is classical; the paper verifies its
specialization to the declared two-traveller lottery model.

{[}8{]} L. de Moura and S. Ullrich. The Lean 4 Theorem Prover and
Programming Language. \emph{Automated Deduction --- CADE 28},
pp.625--635, 2021.
\href{https://doi.org/10.1007/978-3-030-79876-5_37}{DOI}.

{[}9{]} The mathlib Community. The Lean Mathematical Library.
\emph{Certified Programs and Proofs (CPP)}, pp.367--381, 2020.
\href{https://doi.org/10.1145/3372885.3373824}{DOI}. The development
uses the pinned Mathlib 4 version stated above.

{[}10{]} OpenAI. \emph{Sharing AI progress in mathematics}. 6 October
2026.
\href{https://openai.com/index/sharing-ai-progress-in-mathematics/}{Announcement};
\href{https://github.com/openai/math}{mathematical manuscripts and
supporting proof artifacts}. Accessed 20 October 2026. Cited as the
author's motivation for revisiting the thesis, rather than as evidence
for any theorem in this paper.

{[}11{]} OpenAI. \emph{Codex} {[}AI coding and research assistant{]}.
Used October 2026. \href{https://openai.com/codex/}{Product and
documentation}. The precise underlying model snapshot was not recorded;
the resulting research artifacts and validation records are retained in
the companion supplement.

\newpage

\section{Appendix A. Traceable source and proof
correspondence}\label{appendix-a.-traceable-source-and-proof-correspondence}

\begin{longtable}[]{@{}
  >{\raggedright\arraybackslash}p{(\columnwidth - 2\tabcolsep) * \real{0.5000}}
  >{\raggedright\arraybackslash}p{(\columnwidth - 2\tabcolsep) * \real{0.5000}}@{}}
\toprule\noalign{}
\begin{minipage}[b]{\linewidth}\raggedright
Paper result
\end{minipage} & \begin{minipage}[b]{\linewidth}\raggedright
Declarations, source and added scope
\end{minipage} \\
\midrule\noalign{}
\endhead
\bottomrule\noalign{}
\endlastfoot
Theorem 1, TU & \texttt{fixed\_outsider\_core\_iff};
\texttt{fixed\_outsider\_minima\_exist}. Proposition 4.2.1, p.52/PDF72;
explicit reference-dependent outsider rule \\
Theorem 1, NTU & \texttt{strong\_iff\_fixed\_ntu\_core}. NTU description
and core, (3.2)--(3.3), p.41/PDF61; vector opportunities retained \\
Theorem 2 & \texttt{two\_player\_nash\_optimum\_core};
\texttt{two\_player\_nash\_pareto\_strong}. Pairwise bargaining and
Pareto example, pp.55--59/PDF75--79; efficiency condition explicit \\
Actual source instance &
\texttt{ExampleOne.fixed\_outsider\_core\_exists};
\texttt{ExampleOne.fixed\_outsider\_ntu\_core}. Figure 4.2, equations
(4.2)--(4.5), Table 4.1, pp.58--59/PDF78--79 \\
Theorem 3 & \texttt{ResourceModel.minima\_singleton\_dominated};
\texttt{solo\_le\_member}; \texttt{singleton\_total};
\texttt{coalition\_minima\_exist}. Resource costs, §3.2.1
pp.39--42/PDF59--62; §4.5 pp.64--65/PDF84--85; §6.3.2
pp.124--125/PDF144--145; explicit monotone additive subclass \\
Theorem 4 & \texttt{dominated\_core\_iff}. Cost core (4.23), p.65/PDF85;
isolation completion \\
Theorem 5 & \texttt{gap\_subsidy\_sufficient};
\texttt{gap\_subsidy\_necessary}. Explicit financing extension of
compensation idea, p.47/PDF67; not an original subsidy theorem \\
Theorem 6 & \texttt{equal\_surcharge\_core};
\texttt{per\_capita\_gap\_necessary};
\texttt{optimal\_per\_capita\_gap}. Explicit approximate-stability
extension; preserves full budget \\
Theorem 7 & \texttt{contract\_implements\_strong};
\texttt{contract\_implements\_tu};
\texttt{finite\_contract\_core\_exists}. Binding assumption p.48/PDF68;
referee penalties p.133/PDF153; explicit price rule and jurisdiction \\
Enforcement limits &
\texttt{ExampleOne.inefficient\_contract\_enforceable};
\texttt{contract\_hides\_physical\_gain}. Actual Table 4.1;
distinguishes physical and implemented costs \\
Theorem 8 & \texttt{empirical\_expected};
\texttt{external\_regret\_empirical\_cce};
\texttt{zero\_regret\_empirical\_cce}. Repeated-game discussion
pp.127--134/PDF147--154; explicit all-action regret \\
Pure-equilibrium efficiency & \texttt{nash\_social\_efficiency}.
Standard smoothness corollary of the same source equations; actual
smoothness required \\
Resource inefficiency counterinstance &
\texttt{EfficiencyCase.raw\_certificate}; \texttt{bad\_is\_nash};
\texttt{endpoint\_costs}; \texttt{optimum\_is\_optimal};
\texttt{no\_finite\_zero\_optimum\_bound};
\texttt{positive\_optimum\_unbounded}; \texttt{not\_smooth}. New
synthetic ring instance n4/seed2018101, source-grounded overlap
subclass; explicit constant private-cost extension \\
Theorem 9 & \texttt{finite\_time\_efficiency}. Equations (6.5)--(6.13),
pp.132--133/PDF152--153; standard corrected quantified statement \\
Financed isolation proposition & \texttt{rebate\_preserves\_nash};
\texttt{financed\_isolation\_counterpart};
\texttt{ResourceModel.financed\_nash\_core};
\texttt{subsidized\_core\_is\_funded\_core};
\texttt{ExampleOne.financed\_source\_example}. Proposition 4.2.1
p.52/PDF72; explicit efficiency and sponsor-financing completion \\
Figure 1 geometry & \texttt{source\_gain\_constraints};
\texttt{source\_nash\_product\_bound};
\texttt{symmetric\_split\_expected\_costs};
\texttt{symmetric\_split\_attains\_nash\_product}. Table 4.1 p.59/PDF79;
declared status quo and joint-lottery representation; known Nash-product
interpretation \\
Theorem 12 & \texttt{Negotiation.Protocol.run\_safe};
\texttt{rank\_decreases}; \texttt{run\_terminal};
\texttt{negotiate\_correct}; \texttt{checked\_correct};
\texttt{planning\_guarantee}. Algorithms 1--2 pp.55--57/PDF75--77;
explicit sequential scan, acceptance, strict-descent and stopping rules.
Executable definitions, not a proof of an unspecified threat-based
implementation \\
Theorem 11 & \texttt{exists\_approved\_minimum};
\texttt{exists\_baseline\_safe\_plan}. Assumption 4.3.1, p.56/PDF76 and
personalized strategy generation; explicit finite baseline/consent
completion, not a theorem about the original heuristic \\
Theorem 10 & \texttt{transfer\_observed\_regret};
\texttt{distortion\_of\_uniform\_error};
\texttt{efficiency\_with\_observed\_costs}. Personal/system costs
p.81/PDF101 and p.125/PDF145; changing prices p.134/PDF154; fixed
physical game and bounded cumulative distortion \\
\end{longtable}

\end{document}